\documentclass[12pt]{article}

\usepackage{graphicx}

\title{ The colormagnetic confinement in QCD}
\author{  Yu.A.Simonov \\
  NRC ``Kurchatov Institute'' -- ITEP
 \\
Moscow, 117218 Russia}

\newcommand{\beq}{\begin{eqnarray}}
 \newcommand{\eeq}{\end{eqnarray}}
\newcommand{\be}{\begin{equation}}
 \newcommand{\ee}{\end{equation}}

\def\fun#1#2{\lower3.6pt\vbox{\baselineskip0pt\lineskip.9pt
\ialign{$\mathsurround=0pt#1\hfil ##\hfil$\crcr#2\crcr\sim\crcr}}}

\newcommand{{\SD}}{\rm SD}

\newcommand{{\Mc}}{\mathcal{M}}

\newcommand{\vex}{\mbox{\boldmath${\rm x}$}}
\newcommand{\vey}{\mbox{\boldmath${\rm y}$}}
\newcommand{\ver}{\mbox{\boldmath${\rm r}$}}

\newcommand{\vep}{\mbox{\boldmath${\rm p}$}}

\newcommand{\veS}{\mbox{\boldmath${\rm S}$}}
\newcommand{\veL}{\mbox{\boldmath${\rm L}$}}

\newcommand{\veu}{\mbox{\boldmath${\rm u}$}}

\newcommand{\llan}{\langle\langle}
\newcommand{\rran}{\rangle\rangle}
\newcommand{\lan}{\langle}
\newcommand{\ran}{\rangle}

\begin{document}
\maketitle
\begin{abstract}
Colormagnetic confinement as a natural component of the QCD confinement is explained and treated in the
framework of the Field Correlator Method. For quarks and gluons in hadrons the effects of the colormagnetic confinement
are discussed at zero temperature, where it contributes to the spectrum properties and can create its
own bound states, while at nonzero temperature in the EoS  of the quark gluon plasma the colormagnetic confinement plays a dominating role. Its properties in the QCD thermodynamics are discussed in detail. In particular the CM string tension and the Debye screening mass calculated in FCM are compared with lattice data.

 \end{abstract}

\section{Introduction}
The confinement in QCD is a general phenomenon which establishes main features of our Universe, yielding more than 90 percent of its visible mass.
The theory of colormagnetic confinement (CMC) and colorelectric confinement (CEC) based on the Field Correlator Method (FCM) has been formulated in the form of analytical approach \cite{1,2,3,4,5} and studied numerically  by the lattice data \cite{5*,6,7,8,9}, which support the good convergence of the method . Since that time the CEC was studied in detail and its basic mechanism -FCM where correlators of field strength (FS) are calculated selfconsistently via integrals of FS -- was exploited in numerous analysis of experimental and lattice data -see  \cite{5} for recent review. The role of CMC is less evident since at $T=0$ it appears as the spin-dependent corrections in hadron spectra and reactions. Moreover, the CMC also defines the basic interaction of quarks and gluons at high temperature and in the quark-gluon plasma. The CMC is provided by the colormagnetic field correlators in the same way as the standard CEC arises from the colorelectic correlators and at zero temperature both correlators coincide. However, they are yielding completely different contributions to the hadron dynamics: the CEC establishes the main part of the visible hadron mass of the Universe, while  at zero temperature the main role of the CMC is providing one half of the vacuum field energy and establishing  the spin and the momentum-dependent terms in the hadron
Hamiltonian. This provides important corrections in the hadron spectra as will be discussed below.  With increasing temperature the roles of both confining forces change drastically:
the CEC is decreasing and finally disappears at the critical temperature, $T\geq T_c$, while the CMC grows (with the CMC string tension  increasing as $T^2$) and plays an important role in the quark-gluon dynamics. For that reason  the analysis of the quark-gluon plasma requires the account of the CMC.

The important role in the analysis of the CMC was always played by lattice analysis \cite{1a,2a,3a}
which revealed from the very beginning that CMC is not like CEC for temperatures $T>T_c$ and moreover
the colormagnetic string tension $\sigma_s$ at large $T$ is proportional to $T^2$ \cite{2a,3a}. It was understood that CMC could be analyzed in the $3d$ model with the adjoint Higgs field \cite{4a,5a,6a,7a}
Moreover the analysis of the gluon screening mass has allowed the lattice measurement of the nonperturbative Debye mass $m_D(T)$ \cite{8a,9a,10a}.

 We shall demonstrate below the analytic calculation of both CMC string tension $\sigma_s$ and $m_D(T)$ in the framework of FCM and display a good agreement with lattice data. At this point it is important
 to stress that FCM enables one to calculate the field correlators (both CM and CE) as two-gluon Green's functions (gluelumps) $G_{gl}$  where gluons interact via CM and CE confinement and the resulting equation for the string tension is an integral of $G_{gl}$ with gluons interacting via the same $\sigma$. This gives a check of selfconsistency of the whole method and as we shall show below it enables one to calculate $\sigma_s(T)$ without extra parameters in agreement with lattice data.
 Summarizing the additional features of the CMC (being the important part of the
general nonperturbative FCM method), one discovers the strong spin-orbit force (``the Thomas term") in hadron spectroscopy, the strong coupling effect in the qgp, the origin of the effective screening mass at $T>T_c$, the resolution of the Linde problem in the high $T$ perturbation theory. It is the purpose of the present paper to summarize the existing knowledge of the CMC and to
propose possible developments in this field, which can be checked both numerically and experimentally.
For many years the confinement theory was also using different ideas based on the geometrical or quasiclassical objects in the
QCD vacuum, such as monopoles or center vortices (see \cite {11a,12a} and \cite{13a,14a} as  reviews). In principle this can be accomodated in the FCM as an additional (might be unnecessary) detailisation of the FCM correlators, whereas the method can keep its form.
 In this sense one consider this approach as an attempt to understand why at all field strength correlators have nonzero vacuum averages in the confining phase. The present paper gives an answer to this question and predict CM correlators at low and high
temperatures both in the confining and deconfined phase.
 The plan of the paper is as follows. In the next section we introduce the field correlators responsible for the colorelectric (CE) and colormagnetic (CM) confinement and construct the hadron Hamiltonian containing both effects. In section 3 we specifically study the CM effects in hadrons in the phase of the CE confinement. The section 4 is devoted to the CM interactions in the CE deconfined phase where we discuss the analysis of the CM effects in quark-gluon plasma which yield the growing as $T^2$ the CMC string tension. We also analyze the standard perturbative theory in qgp and  using the CMC, we distinguish and resolve the famous Linde problem. The concluding section contains the overall discussion of the results and an outcome.

 \section{The colormagnetic and colorelectric correlators and the QCD Hamiltonian}

The CEC in the framework of FCM was exploited as a basic dynamical theory for the hadron spectra and wave functions \cite{10,11,12,13,14} with numerous applications \cite{15,16,17,18,19,19*,20,20*,20**,21,22,23,24}.
 To give a simple idea of the FCM we can describe the following picture of the hadron in the QCD.
In QCD the quarks and gluons propagate along Wilson lines and the propagation of all hadrons can be described by the corresponding Wilson loops, which  according  to the nonabelian Stokes theorem  contain inside numerous field fluxes which in the certain gauge (``the generalized contour gauge", see \cite{3} for details and discussion) can be written simply as $ F_{\mu,\nu}(z) d\sigma_{\mu,\nu}(z)$, actually, the integral of those.

In the FCM one considers these fluxes, with all ${z}$ inside the Wilson loop, as a statistical medium with the field correlators, defined by the average values of $\lan F(x)F(y>, <F(x)F(y)F(z)\ran, ...$. It was proved that in FCM the lowest correlators $\lan F(x)F(y)\ran$ are dominant, while the higher ones contribute
less than 5 percent in agreement with detailed lattice data \cite{5}. This result refers to the time-like $F_{i4}= E_i$, as well as to the space-like $F_{ik}= e_{ikl} H_{l}$ field strengths.
This stochastic concept, fully supported by existing data, will be the basis of our analysis here, mainly devoted to the CMC, described by the colormagnetic field correlators $\lan H_i(x) H_k(y)\ran$,  and the resulting physical phenomena.

We start with the definition of the field correlators, both colorelectric and colormagnetic.

$$
\frac{g^2}{N_c}\llan TrE_i(z)\Phi(z,z') E_j(z')\Phi(z',z)\rran= $$\be =\delta_{ij} \left[D^E(u) + D^E_1(u) + u^2_4 \frac{\partial D^E_1}{\partial u^2}\right] + u_i u_j \frac{\partial D^E_1}{\partial u^2},
\label{1}\ee
$$
\frac{g^2}{N_c} \llan Tr H_i(z)\Phi(z,z') H_j(z')\Phi(z',z)\rran=$$\be= \delta_{ij} \left[D^H(u) + D^H_1(u) + \veu^2 \frac{\partial D^H_1}{\partial u^2}\right] - u_i u_j \frac{\partial D^H_1}{\partial u^2},
\label{2}\ee

\be\frac{g^2}{N_c} \llan Tr H_i(z)\Phi(z,z') E_j(z')\Phi(z',z)\rran= e_{ijk} u_4 u_k \frac{\partial D^{EH}_1}{\partial u^2}.\label{3}\ee

Here the resulting correlators $D^E(u),D^H(u)$ define the confinement interaction -- the string tensions- in the planes $(i4),(ik)$, namely,
\be
\sigma^{E(H)}= \frac{1}{2} \int d^2 z D^{E(H)}(z).
\label{4}\ee
It is important that at zero temperature all Euclidean planes are equivalent and both colormagnetic (CM) and colorelectric (CE) correlators coincide, as well as the string tensions, and each hadronic system is under the action of both colorelectric  and colormagnetic forces. However, above the critical temperature $T_c$ the colorelectric correlators vanish and the QCD vacuum is fully
in the realm of the CM correlators (apart from the perturbative interactions). It is the purpose of the present paper to study specifically the effects
of the colormagnetic interactions both, below $T_c$ -- the colorelectric confinement region, and above $T_c$ -- in the CMC region.
In this section we derive the Hamiltonian with the CEC and CMC in the quark-antiquark systems.

The Hamiltonian for heavy quarks in terms of the field correlators was written in \cite{10}.
To derive the Hamiltonian in the case of light quarks  one can use the relativistic Fock--Feynman--Scwinger path integral method \cite{11}, which relates the integral representation of the $q\bar q$ Green's function with the Hamiltonian in terms of the virtual quark (antiquark) energies $\omega_1$ ($\omega_2$). Its general form was elaborated in \cite{14,16}. We follow below the form of \cite{15}, where the result is presented in terms of $\sigma_H,\sigma_E$.
The general form of the Hamiltonian consists of the radial kinetic term $H_0$, the orbital motion term $H_l$, the spin-dependent term $H_{sd}$, the perturbative contribution $H_{pert}$ and the self-energy term $H_{se}$. To make the complicated general form of the Hamiltonian more simple it is convenient to introduce the extra parameters (called ``einbeins''), which are defined via
the solution of the subsidiary equations for the resulting energy (mass) eigenvalues,
\be
\frac{\partial M(\lambda_i)}{\partial \lambda_i}=0, \lambda_i= \omega_1, \omega_2, \nu(\beta), \eta.
\label{5} \ee
The Hamiltonian can be written as
\be
H(\omega_1,\omega_2,\nu)= H_0 + H_{int}  + H_l + H_{sd} + H_{pert} + H_{se}.
\label{6} \ee

Here $H_0$ contains only the radial kinetic motion and is written in terms of quark and antiquark
effective energies $\omega_1,\omega_2$. In what follows we shall discuss the case of the equal masses, with correspondingly $\omega_1=
\omega_2= \omega$. The case of general mass relations can be found in \cite{14}.

\be
H_0= \omega + \frac{p^2_r + m^2}{\omega}.
\label{7} \ee

\be
H_{int}= \int^1_0d\beta \left[\frac{\sigma_1^2  r^2}{2\nu} + \frac{\nu}{2} + \sigma_2 r\right].
\label{8} \ee
Here $\sigma_1= \sigma_H + \eta^2(\sigma_H- \sigma_E), \sigma_2= 2 \eta (\sigma_E- \sigma_H)$.
The orbital part of the Hamiltonian, $H_l$ depends not only on the effective energies but also
on the colorelectric and colormagnetic string tensions, expressed via the einbein factor $\nu(\beta)$, see \cite{14,14*},
\be
H_l= \frac{\veL^2}{r^2 [\omega + 2 \int^1_0d\beta (\beta- 1/2)^2 \nu(\beta)]}.
\label{9} \ee.

The most complicated term of the Hamiltonian is the spin-dependent part, derived in \cite{10,19*,20*,20**},
$$
H_{sd}=\left (\frac{\sigma_i^{1}}{4\omega_1^2}+\frac{\sigma_i^{2}}{4\omega_2^2}\right)L_i \frac{1}{r}
(V'_0(r) + 2 V'_1(r)) + \frac{\sigma_i^{1}+ \sigma_i^{2}}{2 r \omega_1 \omega_2} V'_2(r) +$$\be+
\frac{3 \sigma_i^{1} n_i \sigma_k^{2} n_k - \sigma_i^{1} \sigma_i^{2}}{12\omega_1 \omega_2} V_3(r) +
\frac{\sigma_i^{1}\sigma_i^{2}}{12\omega_1 \omega_2} V_4(r).
\label{10}\ee
Here the spin-dependent potentials are expressed via the field correlators $D^E,D^E_1,D^H,D^H_1$
where the last two correlators are appear due to the CMC, namely,
\be
V'_0(r)= 2 \int_0^\infty d\nu \int^r_0 d\lambda D^E(\lambda,\nu) +r\int_0^\infty d\nu D^E_1(r,\nu),
\label{11}\ee
\be
V'_1(r)= -2\int_0^\infty d\nu \int_0^r d\lambda (1- \lambda/r) D^H(\lambda,\nu)
\label{12}\ee
\be
V'_2(r)= \frac{2}{r}\int_0^\infty d\nu \int_0^r\lambda d\lambda D^H(\lambda,\nu) + r\int_0^\infty d\nu D^H_1(r,\nu)
\label{13}\ee
\be
V_3(r)= -2r^2\frac{\partial}{\partial r^2} \int_0^\infty d\nu D^H_1(r,\nu)
\label{14}\ee
\be
V_4(r)= 6\int_0^\infty d\nu \left[D^H(r,\nu)+ (1+ \frac{2r^2}{3}\frac{\partial}{\partial \nu^2}D^H_1(r,\nu))\right]
\label{15}\ee
Here  the field correlators depend on only one variable: $D(x,y)=D(\sqrt(x^2+ y^2)$.
One can see important contribution of the CMC terms, $D^H$ and $D^H_1$, which  define the spin-spin forces, and one may wonder what is the contribution of their purely nonperturbative parts.
To this end we are using (\ref{8})-(\ref{12}) in the large $r$ region  and obtain the estimates,
\be
 V'_0/r= 1/r \sigma^E + O(1/r^2),
\label{16}\ee
\be
 V'_1/r= - 1/r \sigma^H + O(1/r^2),
\label{17}\ee
while $V_3,V_4$ terms decay exponentially at large $r$. As a result at large r one obtains  the dominant
contribution for the spin-orbit force $V_{ls}$ in the case of equal quark and antiquark mass  (the first two terms in (\ref{7}),
\be
V_{ls}= \frac{\veS \veL}{2 \omega^2 r} (\sigma^E- 2\sigma^H) + O(1/r^2).
\label{18}
\ee
This expression can be compared with purely perturbative contributions to the spin-dependent interactions, where the CMC and CEC do not appear, which, however, can be derived from
the correlators $D^{E,H}_1$. To this end we can identify the purely perturbative spin-dependent contributions $V_{ip}(i=1,2,3,4)$, namely \cite{15*},
\be
V_{1p}=0,~~ V'_{2p}= \frac{4\alpha_s}{3r^2}, ~~V_{3p}= \frac{4 \alpha_s}{r^3}, ~~V_{4p}= \frac{32\pi \alpha_s \delta^{3}(\ver)}{3}.
\label{19}\ee
Here we have suppressed the indices $E,H$ in the perturbative expressions. Note that the strong coupling constant $\alpha_s$ is well defined in the coordinate space,
since the QCD constant $\Lambda_{\overline{MS}}(n_f)$ is now known from experiment \cite{24*,24**}.
Finally, we need to take into account the self-energy contribution to the Hamiltonian $H_{se}$, which is a definite
negative constant, produced by the $\sigma_{\mu\nu} F_{\mu\nu}$ part in the Green's function
\cite{25,26},
\be
H_{se}= -\frac{4\sigma_E}{\pi \omega^{0}} \chi(m_q).
\label{20} \ee
where $\omega^{0}$ is the stationary value of the effective quark energy, obtained as in (\ref{5}), and $\chi(m_q)= 0.9$ for the zero quark mass and $\chi= 0.80$ for the $s$ quark.
At this point we stress that the resulting Hamiltonian (\ref{6}) does not contain any fitting constants and is fully defined by the field correlators $D^E,D^E_1,D^H,D^H_1$,
while its spin-independent part is defined by only $\sigma_E,\sigma_H$. This is specifically true for the FCM Hamiltonian, while all other existing approaches exploit numerical fitting constants or functions. In the next chapter we shall discuss the comparison of our theoretical results with experimental and lattice data, making a special emphasis on the role of the CMC contributions.

\section{The colormagnetic interaction in hadrons at zero temperature}

 We start our analysis of the resulting Hamiltonian (\ref{6}) with the spin-independent part and firstly consider  the case of $L=0$. At $T=0$ both string tensions are equal $\sigma_E= \sigma_H$,
giving $H_l=0$ and varying over $\nu,\eta$, one obtains the simple result,
\be
H_{int}= (\sigma_1 + \sigma_2) r = \sigma_E r .
\label{21} \ee
However, taking into account that $H_l$ also contains $\nu$ and therefore should participate in the varying
(optimization) process, and keeping unequal $\sigma_E,\sigma_H$, one obtains approximately \cite{15}
\be
H_{int} + H_l= \eta \sigma_E r + (1/\eta -\eta) \sigma_H r + \omega y^2, \eta=\frac{y}{\arcsin y},        \label{22} \ee
where $y$ is a solution of the equation,
\be
\frac{\sqrt{L(L+1)}}{\sigma_H r^2}= 1/(4y) (1 + \eta^2(1-\sigma_E/\sigma_H))(1/\eta -\sqrt{1-y^2}) + \frac{\omega y}{\sigma_H r}.
\label{23} \ee
One can see in (\ref{23}) that for $\sigma_E=\sigma_H$ and $L=0$ one has $y=0$ and the resulting $H_{int} + H_l= \sigma r$, while for $L > 0$ the presence of the parameter $\nu$ in the denominator of (\ref{9}) (which denotes the string contribution to the rotating mass) brings the so-called string correction in the Hamiltonian. For example, in the heavy quark system this gives
$\Delta H_l= -\frac{\sigma \veL^2}{6m^2 r}$. One can see that $\sigma_H$ plays an important role in the hadron dynamics at $T=0$.

Turning to the spin-dependent dynamics one can write the most important nonperturbative contribution in (\ref{10}); the analysis of the
resulting expressions for the spin-dependent potentials, made in \cite{15*,19*}, shows that the nonperturbative CMC contributions to the tensor and spin-spin forces are strongly suppressed, while the spin-orbit forces are dominated by them. Indeed, writing the spin-orbit term from (\ref{11}),(\ref{12})
and neglecting the terms $D^E_1,D^H_1$,
\be
V_{so}(r)= \left(\frac{\veS_1\veL_1}{2\omega_1^2}- \frac{\veS_2\veL_2}{2\omega_2^2}\right)= 1/r(V'_0 +2V'_1),
\label{24}\ee
then neglecting $D^E_1$, which produces small contribution at low values of $r$, one has
\be
1/r V'_0= 2/r \int^\infty_0 d\tau \int^r_0d\lambda D^E(\tau,\lambda),
2/r V'_1= -4/r \int^\infty_0 d\tau \int^r_0 d\lambda D^H(\tau,\lambda) (1- \lambda/r).
\label{25}\ee

At zero temperature $\sigma^E= \sigma^H$ and due to $D^H$ one obtains the full contents of the famous negative  Thomas term \cite{27},
which was the object of numerous studies, see e.g. \cite{20} for the field correlator treatment and
\cite{28} for the string dynamics approach.

Indeed the phenomenological Thomas term for heavy quarks \cite{27}, $V_{so}= -\frac{\sigma \veS\veL}{2 m^2 r}$, is produced by both the CMC and CEC, connected by the Gromes relation \cite{29} at $T=0$ with $\sigma=\sigma^E=\sigma^H$ (see \cite{10,20**,19*} for more details). In this way one can see that the CMC secures the
correct behavior of the spin-orbit forces in hadrons.
To understand  how it works in reality we can calculate the nonperturbative spin-orbit matrix element
for the $nP$-states with the radial excitation $n$:  $a_{so}(nP)= - \frac{\sigma_H \lan 1/r\ran}{2 \omega^2(nP)}$, where $\omega(nP)$ is the effective quark energy, defined in (\ref{5}). The analytically computed values of $a_{so}(nP)$ for the ground ($n=1$) states  and different $q\bar q$ systems are given below in Table 1.

\begin{table}[!htb]
\caption{Nonperturbative spin-orbit splitting of the $1P$ quark-antiquark states}
\begin{center}
\label{tab.01}
\begin{tabular}{|l|c|c|c|}
\hline
$ q\bar q$                  & $n\bar n$&   $c\bar c$&   $ b\bar b$\\\hline

$ a_{so}^{np}(1P)$ (in MeV) & -88     & -13.3     & -2.3 \\

$<1/r> (1P)$ (in GeV$^{1/3}$)  & 0.241   & 0.394    & 0.559 \\

$r^{-3}$ (in GeV)       &  0.0271     &   0.120    & 0.448 \\

                           & th (exp) & th (exp)  & th (exp) \\

$a_{so}^{tot}(1P)$ (in MeV)  & 41 (abs)  & $34.0 (35.0\pm 0.2)$& $13.3 (13.6\pm 0.7)$\\
\hline
\end{tabular}
\end{center}
\end{table}

Here one can see  strong decrease of the nonperturbative spin-orbit term with the growing quark mass,  which is very small in bottomonium, whereas in a light meson its magnitude is large, providing
decreasing of the fine-structure splitting, in agreement with the experimental data.

It is now interesting to compare our results for glueballs with the lattice data \cite{30}, as it was done in \cite{20**}. For glueballs the total scheme of the spin-dependent forces is the same as for mesons, given above, except that all field correlators and the string tension are $9/4$ times larger. The comparison of the FCM prediction for the states $0^{-+},2^{-+}$, split by the spin-orbit interaction, is as follows \cite{20**} (in GeV): $M_{glb}(0^{-+})= 2.56 , M_{glb}(2^{-+})= 3.03$, which can be compared with the lattice data \cite{31}: $2.59$~GeV and $3.1$~GeV, respectively.
Note that the FCM calculations do not contain any fitting parameters, while the overall negative
constant in the Hamiltonian in (\ref{5}) is calculated via the string tension \cite{25,26}. Also in the FCM there is no fitting parameters for all low-lying mesons \cite{32} and only highest states need corrections due to so-called ``flattening'' of the confinement potential, which occurs  due to holes in the film, produced by the pair creation process \cite{32}. This is in contrast to the well-known calculations of hadron masses \cite{33}, where multiple fitting constants are used and the overall subtraction constant is introduced. We would like to underline that in the FCM the negative  correction $H_{se}$ is calculated via string tension and the quark kinetic energies.

Summarizing one can say that the CMC defines the important part of the strong spin-orbit interaction
in hadrons at zero temperature, while the the CEC defines the linear confinement interaction, and the perturbative QCD is mostly responsible for
the short-range spin-spin forces.

\section{The CMC at finite temperature and in the quark-gluon plasma}

One can consider the $T>0$ region in two aspects:\begin{enumerate}
 \item as an individual hadron physics in the regions with $\sigma_E < \sigma_H$  and in the deconfined region $\sigma E=0$, \item the role of the CMC in the
thermodynamics of quark-gluon plasma (qgp). Below we discuss these points in this order.
\end{enumerate}

1. ~  It was shown in \cite{15} that the resulting spin-orbit potential (\ref{24}),(\ref{25}) has the
form of the attractive Thomas potential at large $r$ and strong repulsive core at small distances, which ensures weakly coupled $q\bar q$ bound states for the quark mass $m_q>0.22$ GeV
(due to the CMC contribution); e.g. for $s$-quark with $m_s=0.22$~GeV  the $s\bar s$ binding energy is $-45$ MeV and much less for the $c$ and $b$-quarks. The situation for light quarks
in the deconfined region is even more complicated and seems to be similar to the $Z>g 137$ critical phenomena in QED, when the central charge Z is surrounded by the plasma-like vacuum \cite{15}. In difficult to develop the quantitative theory of the
corresponding medium at the deconfining temperatures around $150$ MeV  but one can expect that
these effects will give relatively small corrections at this temperature.

2.~   We shall turn now to the most important topic of the role of the CMC in the quark-gluon thermodynamics at
$T>T_c$ and show that the CMC will provide the following basic features in this region:

A) the growth of $\sigma_H(T)$ with the temperature, $\sigma_H(T)={\rm  const}~ T^2$;
B) the effects of the CMC on the quark-gluon medium which gives a special CMC factor
in the pressure of quarks and gluons;
C) the mass correlation parameter (the Debye mass) defining
the gluon exchange forces in qgp in the background of CMC vacuum;
D) the violation of the standard perturbation theory in the qgp,  when the $g^6$ term contains the infinite series of contributions- the Linde problem. We shall below discuss these topics term by term.

 \subsection {(A). The colormagnetic string tension $\sigma_H$ at nonzero temperature.}
As was discussed in the Introduction this topic was actively studied on the lattice \cite{1a,2a,3a,4a} where also the model
containing an adjoint Higgs was exploited with similar results \cite{5a,6a,7a}. On the theoretical side one can express
in the framework of FCM the CM string tension via the gluelump Green's function, where gluelump is the system of 2 gluons and
adjoint Wilson line connected by adjoint strings. Actually gluelumps define the confining dynamics in both CE and CM strings
in a selfconsistent way since CE and CM string tensions are expressed via integrals of the corresponding gluelump Green's functions,where interaction is given again by the CE and CM string tensions. The behavior of $\sigma_H$ near $T_c$ was found
 in \cite{34} in good agreement with lattice data. In the large $T$ region  the FCM allows to define it analytically  \cite{35}  and compare with lattice data \cite{36} in \cite{36*} finding a good agreement.
We shall be interested in the region of temperatures $T>T_c$ and exploit the standard definition
\ref{4} of $\sigma_H$ via the CM correlator,
$\sigma_H= 1/2 \int d^2z D^H(z)$, where $D^H$ is expressed via two-gluon Green's function and finally via the product of interacting $4d$ one-gluon Green's functions $D^H(z)\sim G^{(2g)}_{4d}(z)\sim(G^{(g)}_{4d})^2$. It is important \cite{34,35} that the path integral along the 4-th axis does not contain interaction and therefore at large $T$ one arrives at the result
\be
G^{g}_{4d}= T G^{(g)}_{3d}(z) + K_{3d}(z): ~~ D^H(z)= \frac{g^4 (N^2_c -1)T^2}{2} \lan G^{2g}_{3d}(z)\ran + ...,
\label{26}\ee
where neglected terms are subleading at large $T$.
As a result the CMC string tension at large $T$ can be written as
\be
\sqrt{\sigma_H(T)}= g^2 T c_{\sigma} + {\rm const}~, ~~c^2_{\sigma}= \frac{N^2_c-1}{4}\int d^2 w \lan G^{2g}_{3d}(w)\ran
\label{27}\ee
Numerically the lattice data \cite{36} yield $c_{\sigma}= 0.566 +/- 0.013$.
In FCM using (\ref{27}) the integral was calculated approximately yielding as a lower limit $c_{\sigma}= 0.47$ in a reasonable agreement with lattice.
We now turn to the  region $0<T<T_c$, where one can generalize the form of $D^H(z)$ to the nonzero $T$ region, summing over infinite $n\beta$ series ($\beta= 1/T)$ \cite{34}, which yields at small $T$
\be
\sigma_H(T)/\sigma_H(0)= \frac{\sin h(M/T)+ M/T}{\cosh(M/T)- 1}= 1+ 2(1+M/T)\exp{-M/T}+ O(\exp -2M/T).
\label{28}\ee

\subsection{(B). The CMC pressure in the quark-gluon plasma}

Using the relativistic path integral for the quark and gluon pressure $P_q,P_g$ \cite{37,13}, one express those via the spatial loop integrals of the thermal Green's functions of $q,g$ respectively,  $G_3(s),S_3(s)$
\be
P_{gl}= \frac{N^2_c-1}{\sqrt{4\pi}}\int^\infty_0\frac{ds}{s^{3/2}} G_3(s) \sum{n}\exp(-\frac{n^2}{4T^2 s}) L^{n}_{adj},
\label{29}\ee
where $L^{n}_{adj}$ is the adjoint Polyakov loop  and $G_3(s)$ is the 3d closed loop gluon Green's function as a function of the relativistic square of distance $s$. It is clear that in the 3d closed loop the confinement is colormagnetic and the result for q and g Green's functions can be written as \cite{38}
\be
G_3(s)= \frac{1}{(4\pi s)^{3/2}} \left(\frac{(M_{adj})^2 s}{sh((M_{adj})^2 s)}\right)^{1/2}
\label{30}
\ee
Here $M_{adj}= 12 \sqrt{\sigma_H(T)}$. For the quark function $S_3(s)$ one should replace $M_{adj}$ by $M_f= 1/3 M_{adj}$.
Substituting (\ref{30}) into (\ref{29}) one obtains the gluon pressure as
\be
P_{gl}= \frac{2(N^2_c-1)}{(4\pi)^2}\sum_{n=\infty,n=1}L^{n}_{adj} \int^{\infty}_0 ds \frac{1}{s^3}
\exp\left(-\frac{n^2}{4T^2s}\right) \sqrt{\frac{M^2_{adj} s}{sh(M^2_{adj}s)}}.
\label{31}\ee
Here $L^{n}_{adj}$ can be taken from lattice \cite{39}  or analytic \cite{38} expressions.

{
\begin{figure}[htb] 
\setlength{\unitlength}{1.0cm}
\centering
\begin{picture}(6.8,6.8)
\put(0.5,0.5){\includegraphics[height=5.7cm]{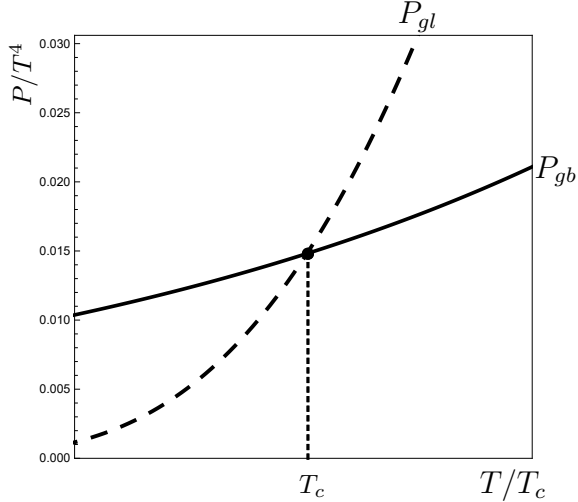}}
\put(6.35,0.1){$T/T_c$}
\put(3.95,0.1){\footnotesize $T_c$}
\put(0.1,5.25){\footnotesize \rotatebox{90}{$P/T^4$}}
\put(5.265,6.35){$P_{gl}$}
\put(7.08,4.3){$P_{gb}$}
\end{picture}
\caption{Pressure $P(T)$ as function of temperature $T$ for the confined phase (Glueballs) -- solid line, and for the deconfined phase (dashed line). The intersection point is at the critical temperature $T_{c}$.}
\label{fig:fig01}
\end{figure}
}\medskip


In Fig. 1 we show how  proceeds the transition of the confined phase of glueballs into the deconfined phase of gluons with the CMC interaction in the gluon plasma in comparison with the lattice data from  \cite{40} (Fig. 1 from \cite{38}).

{
\begin{figure}[!htb] 
\setlength{\unitlength}{1.0cm}
\centering
\begin{picture}(8.0,4.1)
\put(0.5,0.5){\includegraphics[height=3.65cm]{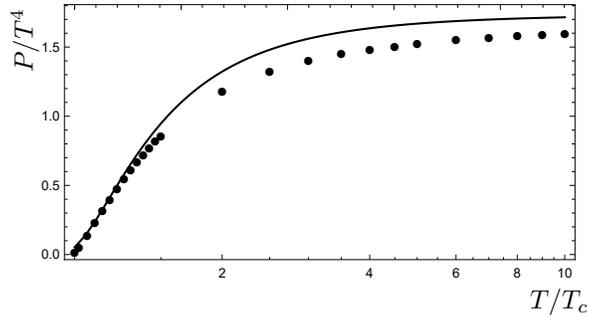}}
\put(7.0,0.1){\footnotesize $T/T_c$}
\put(0.1,3.25){\footnotesize \rotatebox{90}{$P/T^4$}}
\end{picture}
\caption{The pressure ${P(T)}/{T^4}$ in the $SU(3)$ theory in the deconfined phase. The solid line is for the modified oscillator confinement Eq. (\ref{30}), and filled dots are for the lattice data  \cite{40}.}
\label{fig:fig04}
\end{figure}
}\medskip

{
}\medskip

In Fig. 2 we demonstrate the behavior of the gluon plasma at large $T$ vs the lattice data from \cite{40}  (Fig. 3 from \cite{38}).

\subsection{ (C). The Debye mass in the qgp}

In this section we shall show that the only gauge- invariant definition of the Debye mass in the qgp is via the CM mass, i.e. via the square root of the CMC string tension $\sigma_H$, and we shall demonstrate a good agreement between the resulting theoretical and lattice data \cite{41,42}. The problem of the Debye mass in the QCD standard perturbation theory (SPT) is that it cannot be defined in a gauge-invariant way and therefore one is using some approximate definitions, introducing fitting constants, e.g. in \cite{36} the ansatz was exploited
$m_D(T)= A g(T) T \sqrt{1+ N_f/6}$  with $A= 1.51,1.42$ for $N_f= 0,2$, respectively.
Instead in the nonperturbative FCM one can calculate Debye mass with a good accuracy \cite{41,42}. To this end one defines the Debye mass from the gluon-exchange diagram between trajectories of two charges, see Fig 1 from \cite{42} . It is clear that the gluon distorts the Wilson loop surface of two charges and this additional piece (its 3$d$ projection)
contributes being multiplied with $\sigma_H$ to the gluon action.
In this way one understands that the exchanged gluon, together with its projection on the unperturbed plane of two charges, forms the gluelump \cite{43,44} -- the system of one gluon plus another static with infinite mass. The $T$-dependent Hamiltonian for the gluelumps was derived in \cite{41,42} as

\be
H_n= \sqrt{\vep_{perp}^2 + (2\pi n T)^2}+ \sigma_H^{adj} r_{perp}, n= 0,1,2,...
\label{32}\ee
The corresponding gluelump screening mass spectrum was found in \cite{41}.
The lowest eigenvalue of $H_0 $ is equal to $\epsilon_0= 2.82 \sqrt{\sigma_H}$.
In the next approximation one should take into account the OGE interaction in the gluelump which
yields $\Delta \epsilon_0= -5.08 \alpha_s^{eff} \sqrt{\sigma_H}$. As a result one obtains for the
Debye mass
\be
m_D= \epsilon_0 + \Delta \epsilon_0= 2.06 \sqrt{\sigma_H}
\label{33}\ee

\subsection{(D).  CMC in perturbative  thermodynamics of QCD}.

In this approach one of the problems is the resummation of the infinite series of infrared divergent
gluon-loop diagrams, which are known as hard thermal loops (HTL) resummation.
The perturbation theory of the qgp or purely gluon plasma at $T>T_c$ operates with amplitudes $A_n$ corresponding to diagrams with $n$ vertices, which are produced by the term in the
Lagrangian $L_3= g \partial_\mu a_\nu  f^{abc} a^b_\mu a^c_\nu$. There are numerous studies in this field, see e.g.
\cite{45,46} and a recent review \cite{47}. In this perturbative approach
one does not exploit the notion of the CMC in the deconfined phase of QCD, probably, not realizing
that the deconfined phase implies the absence of the CEC but not CMC. Instead, one can introduce
in this area the notion of the ``magnetic mass" of the gluon to prevent the basic divergencies of the theory without CMC. The latter were designated by Linde \cite{48} and are known as the ``Linde problems". The resolution of these problems with account of the CMC was given in
\cite{35} and can be described shortly below as follows.
The main problem perturbative QCD thermodynamics (PQCDTh), which essentially operates in the 3$d$ space, is the IR or large distance divergence, since the gluon propagator $G(x,y) \sim \frac{T}{|\vex-\vey|}$ is a slowly decreasing function at large distances $X$. Correspondingly, the $n$-th order amplitude behaves as $A_n \sim g^n T^{n/2 +1} X^{n/2-3}$. One can see that the diagrams
with $n >g 6$ gluon vertices diverge at large $X$ -- this is just the Linde problem 1.
One can see that the CMC easily solves this problem. Indeed in 3$d$ the Wilson loops, which cover all the diagram surface, obey the screening law: $W(C)= \exp(-\sigma_H S_{\min})$, where $S_{\min}$ is the
area of the surface, and $S_{\min} \sim X^2$. As a result the amplitude acquires the form
\be
A^{\rm conf}_n= g^n T^{n/2+ 1} \int (dX)^{n/2- 3} \exp(-\sigma_H X^2) \sim g^n T^{n/2+1} (\sqrt{\sigma_H})^{-(n/2-3)} {\rm const}.
\label{34}\ee
Now taking into account \ref{27}, $\sqrt{\sigma_H}= g^2 T {\rm const}$, one comes to the conclusion that
all diagrams with $ n> 6$ yield $A^{conf}_n= g^6 T^4 c_n$ (the Linde problem 2).
As a result one should sum up all the diagrams with $n> 6$, as it is shown in \cite{35}, which are
made finite due to the CMC.

\section{Conclusions}

We have considered  the basic picture of the confined and deconfined matter which is well described  in terms of the colorelectric and colormagnetic field correlators. The latter are obtained
selfcosistently from the nonperturbative QCD vacuum with the basic characteristics -- the gluon condensate, $ G_2= \frac{\alpha_s}{\pi} \lan(F_{\mu \nu})^2\ran$ , which can be taken at the standard value,  $ G_2= 0.012$ GeV$^4$ \cite{49}.
As it was shown in \cite{50}, $G_2$ defines  confinement characteristics in the confined
phase (CEC and CMC) with the energy density \cite{49}
$\epsilon_{\rm vac}= - \frac{11-2/3 n_f}{32} G_2$  and the energy density in the deconfined (only the CMC) area is $1/2 \epsilon_{\rm vac}$. The corresponding pressure $P= -F$ in the confined phase can be written as $P({\rm conf})= |\epsilon_{\rm vac}|+ T^4 p_{\rm hadr}$ and the pressure in the deconfined region is $P({\rm deconf})= 1/2 |\epsilon_{\rm vac}| + T^4 (p_q + p_g)$. Now from the relation $P_{\rm conf}(T_c)= P_{\rm deconf}(T_c)$ one obtains the equation for the transition temperature $T_c$ \cite{13,3} via standard expressions of $p_h,p_q,p_g$ (with or without additional interactions, which will induce small corrections in the $T_c$ values since $\epsilon_{\rm vac}$ is a dominant magnitude),
\be
T_c= \left( \frac{1/2 |\epsilon_{\rm vac}| + p_{\rm hadr}}{p_q + p_g}\right)^{1/4}
\label{35}\ee
As a result (taking free the quark, gluon, hadron pressures),  one obtains in \cite{13,3}  $T_c= 240,150,134$ MeV for $n_f= 0,2,4$,
which is very close to the lattice data $T_c=240,146,131$ MeV .
One can see that $\epsilon(CMC)= 1/2 \epsilon_{\rm vac}$ plays the main role in the definition of
the deconfined phase transition.
As it was shown above, the role of the CMC is even more important. Namely, as it was shown above
in the section 3, in the confined region the CMC ensures an important part of the interaction,  (1) without CMC the sign of the nonperturbative part of spin-orbit force (the Thomas term) would have the opposite sign (see (\ref{18},\ref{25})),(2) The CMC yields the important string correction
 $\Delta H= - \frac{\sigma \veL^2}{6 m^2}$.
 We also discussed the possibility of the weakly bound hadrons due to CMC above $T_c$ (in section 3). Finally in the deconfined region the CMC ensures 3 major effects of the qgp physics:

 ~~I. The CMC creates its own factor $G_3(s)$, \ref{30} in the qgp pressure, \ref{31}, which is
 the main contribution (along with the Polyakov line) to the QCD thermodynamics, which is supported by the lattice calculations (Figs. 1, 2 in section 4).

~~ II. The CMC ($\sigma_H(T)= {\rm const} T^2$) creates the Debye mass $m_D= 2.06 \sqrt{\sigma_H}$ (\ref{33}). Finally, the CMC solves the Linde problem \cite{35} which allows to summarize the
 infinite set of graphs and make the total sum finite.

The author is greatly indebted to A. M. Badalian for advices and contributions in section 3
of the paper, and to N. P. Igumnova for help in preparing the manuscript.

\end{document}